%% file: constrainedRL.tex
\documentclass[conference,letterpaper]{IEEEtran}

\usepackage{amsmath}
\usepackage{amsthm}
\usepackage{amssymb}
\usepackage{bm}
\usepackage{xspace}
\usepackage{xcolor}
\usepackage{graphicx}
\usepackage{psfrag}
\usepackage{url}
\usepackage{framed}
\usepackage{cite}

\newtheorem{theorem}{Theorem}
\newtheorem{proposition}{Proposition}

\theoremstyle{definition}
\newtheorem{definition}{Definition}

\newcommand{\stra}{\mathtt{a}}
\newcommand{\strb}{\mathtt{b}}
\newcommand{\strzero}{\mathtt{0}}
\newcommand{\strone}{\mathtt{1}}

\DeclareMathOperator{\expop}{E}
\DeclareMathOperator{\gf}{G}
\DeclareMathOperator{\repart}{Re}
\DeclareMathOperator{\catop}{cat}

\DeclareMathOperator{\probop}{Prob}

\DeclareMathOperator{\entop}{H}

\title{Deriving the Probabilistic Capacity of General Run-Length Sets Using Generating Functions}

\IEEEoverridecommandlockouts

\author{\IEEEauthorblockN{G. B\"ocherer,
R. Mathar}
\IEEEauthorblockA{Institute for Theoretical Information
Technology\\
RWTH Aachen University\\
52056 Aachen, Germany\\ Email: \{boecherer,mathar\}@ti.rwth-aachen.de}
\and
\IEEEauthorblockN{V.C. da Rocha Jr.,
C. Pimentel}
\IEEEauthorblockA{
Communications Research Group - CODEC\\
Department of Electronics and Systems, P.O. Box 7800\\
Federal University of Pernambuco\\
50711-970 Recife PE, Brazil\\
E-mail: \{vcr,cecilio\}@ufpe.br
}
\thanks{This work has been supported by the UMIC Research Centre, RWTH
Aachen University. This work has also been supported by the Brazilian National Council for Scientific and Technological Development (CNPq).}
}

\begin{document}
\maketitle
\input{abstract}

\input{introduction}

\input{framework}

\input{constrained}

\input{generating}

\input{combinatorial}

\input{probabilistic}

\input{discussion}

\input{conclusions}

%\appendices

%\input{appendix}

\bibliographystyle{IEEEtran}
\bibliography{IEEEabrv,confs-jrnls,Literatur}

\end{document}

%% file: abstract.tex
\begin{abstract}
In \emph{Reliable
Communication in the Absence of a Common Clock} (Yeung \emph{et al.},
2009), the authors introduce general run-length sets,
which form a class of constrained systems that permit run-lengths
from a countably infinite set. For a particular definition of
probabilistic capacity, they show that probabilistic capacity is equal to
combinatorial capacity. In the present work, it is shown that the same result also holds for Shannon's original definition of probabilistic capacity. The derivation presented here is
based on generating functions of constrained systems as developed in \emph{On the Capacity of Constrained Systems} (B\"ocherer \emph{et al.}, 2010) and provides a unified information-theoretic treatment of general run-length sets. 
\end{abstract}

%% file: introduction.tex
\section{Introduction}

A constrained system allows the noiseless transmission of input sequences of weighted symbols that fulfill certain constraints on the symbol constellations. The weight of a symbol can have different practical meanings, e.g., in the context
of magnetic recording systems, ``weight'' usually refers to
``tape-length''; other meanings like ``time'' or ``energy'' are possible, 
depending on the modeled system.

For the design of encoders for such systems, it is of interest to determine the maximum entropy rate at which a random process can generate strings that fulfill the constraints. This rate is called the \emph{probabilistic capacity} and a process that reaches this rate is called \emph{maxentropic}. The probabilistic capacity is equal to the combinatorial capacity in the case that the constraints form a regular language. This was originally shown in \cite{Shannon1948}. In \cite{Khandekar2000}, the authors show this property for a slightly generalized setup, since they allow non-integer valued symbol weights, as long as the set of weights is not too dense. The precise definition of ``not too dense'' is stated in Section~\ref{sec:framework}. 

Recently, in \cite{Yeung2009} the authors introduced the continuous
time asynchronous channel as a model for time jitter in a
communication system with no common clock between the
transmitter and the receiver. As a constrained system, such a channel is defined by an in general countably infinite set of distinguishable run-lengths $\mathcal{W}$ and a finite set of labels $\mathcal{L}$. A run is a substring during which the label does not change. The channel then allows the noiseless transmission of strings where each run has a length in $\mathcal{W}$ and a label in $\mathcal{L}$ and where in addition two subsequent runs always have different labels. We will refer to such channels by \emph{general run-length sets}. Formally, general run-length sets are different from constrained systems that have been considered so far in two ways: first, the set of strings that are accepted do in general not form a regular language, and second, the set of distinct string-lengths does in general not fulfill the ``not too dense'' restriction. As a result, general run-length sets cannot be treated using the framework of finite state machines \cite{Khandekar2000}.  

In \cite{Bocherer2010a}, we presented a new framework for constrained systems based on generating functions. Differing from \cite{Khandekar2000}, this framework is neither restricted to regular constraints nor does it impose the ``not too dense'' restriction to the set of string-lengths. The key result of \cite{Bocherer2010a} is that the entropy rate of input processes is in general upper-bounded by the combinatorial capacity of the system. This result allows us to derive the equality of probabilistic and combinatorial capacity for general run-length sets in a unified manner: we first derive the generating function of general run-length sets, we then use the generating function to calculate the combinatorial capacity, and we finally define an input process whose entropy rate is equal to the combinatorial capacity and thus achieves the maximum. In exactly the same way we derived in \cite{Bocherer2010a} as an illustrating example the equality of probabilistic and combinatorial capacity for $(j,k)$ constraints. 

For a particular definition of probabilistic capacity, the equality of probabilistic and combinatorial capacity of general run-length sets was derived in \cite{Yeung2009}. The definition of probabilistic capacity in \cite{Yeung2009} is based on a notion of entropy rate that is different from Shannon's in \cite[Appendix 4]{Shannon1948}. Because of the different definitions of probabilistic capacity, the derivations presented in our work strongly differ from the derivations in \cite{Yeung2009}. We discuss the relation of our results to the results in \cite{Yeung2009} in Subsection~\ref{sec:discussion}.

The remainder of this work is organized as follows. In Section~\ref{sec:framework}, we state the results from \cite{Bocherer2010a} that we need in this work. In Section~\ref{sec:constrained}, we formally define general run-length sets. In Section~\ref{sec:generating}, \ref{sec:combinatorial}, and \ref{sec:probabilistic}, we derive for general run-length sets the generating function, the combinatorial capacity, and the probabilistic capacity, respectively.

%% file: framework.tex
\section{Constrained Systems, Generating Functions, and Capacity}
\label{sec:framework}

In this section, we shortly provide the main definitions and results from our work \cite{Bocherer2010a}. These form a mathematical framework based on generating functions for the information theoretic treatment of constrained systems. Within this framework, we will derive the probabilistic capacity of general run-length sets in the remaining sections. A detailed discussion and proofs of the theorems can be found in \cite{Bocherer2010a}.
\subsection{Constrained Systems and Generating Functions}
\begin{definition}
\label{def:constrainedSystem}
  A \emph{constrained system} $(\mathcal{A},w)$ consists of a countable set $\mathcal{A}$ of strings
  accepted by the system and an associated weight function $w\colon
  \mathcal{A}\rightarrow\mathbb{R}_{>0}$ ($\mathbb{R}_{>0}$ denotes the positive real numbers) with the following property: if $\stra,\strb\in\mathcal{A}$ and $\catop(\stra,\strb)\in\mathcal{A}$ then $w[\catop(\stra,\strb)]=w(\stra)+w(\strb)$.
\end{definition}
Here and hereafter, $\catop(\stra,\strb)$ denotes the concatenation of $\stra$ and $\strb$.
\begin{definition}\label{def:generatingFunction}
  Let $(\mathcal{A},w)$ represent a constrained system. We define the \emph{generating
    function} of $\mathcal{A}$ by
  \begin{align}
    \gf_\mathcal{A}(s)&:=\sum\limits_{a\in \mathcal{A}}e^{-w(a)s},\qquad s\in\mathbb{C}\label{eq:generatingFunction2}
  \end{align}
where $\mathbb{C}$ denotes the set of complex numbers.
\end{definition}
Let  $\Omega=\{w(a)|a\in\mathcal{A}\}$ denote the set of distinct string weights of elements in $\mathcal{A}$. We
order and index the set $\Omega$ such that
$\Omega=\lbrace \nu_i\rbrace_{i=1}^{\infty}$ with
\mbox{$\nu_1<\nu_2<\dotsb$}. For every $\nu_i\in\Omega$,
$N(\nu_i)$ denotes the number of distinct strings of weight $
\nu_i$ in $\mathcal{A}$. A compact representation of the sequence $\{N(\nu_i)\}_{i=1}^\infty$ can be obtained by rewriting the generating function as
\begin{align}
    \gf_\mathcal{A}(s)&=\sum\limits_{i=1}^\infty N(\nu_i)e^{-\nu_i s}.\label{eq:generatingFunction1}
\end{align}
The growth of the sequence $\{N(\nu_i)\}_{i=1}^\infty$ is determined by the analytic properties of $\gf_\mathcal{A}$ \cite{Flajolet2008}. This property lies at the heart of the results presented in this section.
\subsection{Combinatorial Capacity}
From an information-theoretic perspective, of main interest is the \emph{exponential} growth of the sequence $\{N(\nu_i)\}_{i=1}^\infty$, which is captured by the notion of ``combinatorial capacity''.
\begin{definition} 
\label{def:combinatorialCapacity}
We define the \emph{combinatorial capacity} by
\begin{align}
\mathsf{C}&:=\limsup_{k\rightarrow\infty}\dfrac{\ln\bigl[\sum\limits_{i=1}^k N(\nu_i)\bigr]}{\nu_k}\label{eq:combinatorialCapacity}
\end{align}
\end{definition}
Here and hereafter, $\ln$ denotes the natural logarithm.
Shannon's original definition of combinatorial capacity was
\begin{align}
\mathsf{C}_0&:=\limsup_{i\rightarrow\infty}\dfrac{\ln\bigl[ N(\nu_i)\bigr]}{\nu_i}.
\end{align}
The original definition is only meaningful when the set of distinct string weights $\Omega$ is 
\emph{not too dense}, that is, when there exists some constant $L\geq 0$ and some constant $K\geq 0$ such that for any integer $n\geq 0$
\begin{align}
  \max_{\nu_k<n} k\leq Ln^K\label{eq:notTooDense}.
\end{align}
See \cite{Khandekar2000} and \cite{Bocherer2007} for detailed discussions of the ``not too dense'' property. 
The following theorem shows how the combinatorial capacity of a constrained system is related to its generating function and it further shows that our definition of combinatorial capacity is consistent with the original one.  
\begin{theorem}\label{theo:combinatorialCapacity}
Let $(\mathcal{A},w)$ be a constrained system with the generating function
 $\gf_\mathcal{A}(s)$. 
The following holds:
\begin{enumerate}
\item The combinatorial capacity $\mathsf{C}$ is equal to the abscissa of convergence $Q$ of $\gf_\mathcal{A}$, i.e., $\mathsf{C}=Q$.
\item If the set of distinct string weights $\Omega$ is not too dense, then $\mathsf{C}_0=Q$ and in particular $\mathsf{C}_0=\mathsf{C}$.
\end{enumerate}
\end{theorem}

\subsection{Entropy Rate of Input Processes}
In consistency with \cite[Appendix 4]{Shannon1948}, we define the entropy rate of weighted random processes.
\begin{definition}
\label{def:entropyRate}
The \emph{entropy rate} of a random process \text{$Y=\{Y_i\}_{i=1}^\infty$}, $Y_i\in\mathcal{Y}$ with an associated weight function $w\colon\mathcal{Y}\rightarrow\mathbb{R}_{>0}$ is defined as
\begin{align}
\bar{\entop}(Y):=\limsup_{k\rightarrow\infty}\frac{\entop(Y_1,\dotsc,Y_k)}{\expop[w(Y_1)+\dotsb +w(Y_k)]}.\label{eq:entropyRateProcess}
\end{align}
\end{definition}
The operators $\expop,\entop$ are defined as follows: For a discrete random variable $X$ and a deterministic function $f$, $\expop[f(X)]$ denotes the expected value of $f(X)$ with respect to the probability mass function (PMF) of $X$, and $\entop(X):=\expop[-\ln(X)]$ denotes the entropy of $X$.

Let $Y=\{Y_i\}_{i=1}^\infty$ denote a random process that generates input for a constrained system. Transmitted over the system are the strings $\catop(Y_1,\dotsc,Y_k)$. To ensure that the entropy rate as defined in Definition~\ref{def:entropyRate} actually reflects the entropy rate of the transmitted strings, we have to ensure that $Y$ generates the strings unambiguously, see \cite[Section V]{Bocherer2010a}. The following definition guarantees this.
\begin{definition}
\label{def:inputProcess}
Let $Y=\{Y_i\}_{i=1}^\infty$, $Y_i\in\mathcal{Y}$ be a random process and let $p_k$ denote the PMF of $(Y_1,\dotsc,Y_k)$. Define the sequence of auxiliary random variables $X_k=\catop(Y_1,\dotsc,Y_k)$ with the supports truncated to
\begin{align}
\mathcal{X}_k\!=\!\left\{\catop(y_1,\dotsc,y_k)|(y_1,\dotsc,y_k)\!\in\!\mathcal{Y}^k\!\colon\;p_k(y_1,\dotsc,y_k)\!>\!0
\right\}\nonumber
\end{align}
where $\mathcal{Y}^k$ denotes the Cartesian product of $k$ copies of $\mathcal{Y}$. The process $Y$ is an \emph{input process} of the constrained system $(\mathcal{A},w)$ if the supports $\mathcal{X}_k$
fulfill both of the following conditions:
\begin{enumerate}
\item $\bigcup_{k=1}^\infty\mathcal{X}_k \subseteq \mathcal{A}$.\label{enum:input1}
\item if $j\neq k$, then $\mathcal{X}_j\cap\mathcal{X}_k=\emptyset$.\label{enum:input2}
\end{enumerate}
\end{definition}
Condition 1) ensures that $Y$ generates valid strings and condition 2) ensures that $Y$ does so unambiguously. For a discussion of these conditions in the context of general run-length sets, see Subsection~\ref{subsec:preliminaries} in this work. The entropy rate of an input process of a constrained system relates to the combinatorial capacity as follows.
\begin{theorem}
\label{theo:probabilisticCapacity}
Let $(\mathcal{A},w)$ denote a constrained system. The entropy rate of an input process $Y$ of $(\mathcal{A},w)$ is upper-bounded by the abscissa of convergence $Q$ of $\gf_\mathcal{A}$, and, in particular, it is upper bounded by the combinatorial capacity $\mathsf{C}$ of $(\mathcal{A},w)$.
\end{theorem}
Having defined entropy rate in Definition~\ref{def:entropyRate} and input processes in Definition~\ref{def:inputProcess}, we implicitly defined the probabilistic capacity of a constrained system: it is given by the maximum entropy rate an input process can have.

%% file: constrained.tex
\section{Setup}
\label{sec:constrained}
\subsection{Definition of $\langle\mathcal{W},\mathcal{L}\rangle$}
The class of constrained systems that we consider in this work can be specified by a set $\mathcal{W}$ of \emph{run-lengths} and a set $\mathcal{L}$ of \emph{labels}. The set of run-lengths $\mathcal{W}$ is a non-empty, countable subset of the positive real numbers $\mathbb{R}_{>0}$. The set of labels $\mathcal{L}$ is nonempty and finite. One \emph{run} is the substring of a string during which the label does not change. We refer to the length of a run $r$ by its weight $w(r)$ and we refer to the label of a run $r$ by the label function $l(r)$. The set of allowed strings of such a system is given by
\begin{align}
\langle\mathcal{W},\mathcal{L}\rangle:=\bigl\{&\catop(r_1,\dotsc,r_k)\;\bigl\lvert\bigr.\;k\in\mathbb{N},\nonumber\\
&\text{for } i=1,\dotsc,k\colon w(r_i)\in\mathcal{W} \text{ and }l(r_{i})\in\mathcal{L},\nonumber\\
&\text{for } i=1,\dotsc,k-1\colon l(r_{i+1})\neq l(r_i)\bigr\}
\end{align}
where $\mathbb{N}=\{1,2,\dotsc\}$ denotes the natural numbers.
From this definition, we see that if the cardinality of $\mathcal{L}$ is equal to one, then each string in $\langle\mathcal{W},\mathcal{L}\rangle$ consists of only one run with its length in $\mathcal{W}$ and with its label equal to the unique label from $\mathcal{L}$. From now on, we therefore assume $|\mathcal{L}|\geq 2$. We do \emph{not} require that the set $\langle\mathcal{W},\mathcal{L}\rangle$ fulfills the ``not too dense'' property \eqref{eq:notTooDense}. However, we assume that the limit in \eqref{eq:combinatorialCapacity} exists, i.e., that $\langle\mathcal{W},\mathcal{L}\rangle$ has a well-defined combinatorial capacity.
\subsection{$\langle\mathcal{W},\mathcal{L}\rangle$ is in General not Regular}
It is important to note that $\langle\mathcal{W},\mathcal{L}\rangle$, in general, does not form a regular language. If it would, it could be analyzed within the framework of finite state machines as defined in \cite{Khandekar2000}. Consider as an example $\langle\mathcal{V},\mathcal{K}\rangle$ where
\begin{align}
\mathcal{V} &= \{2^k\vert k\in\mathbb{N}\}\\
\mathcal{K} &= \{\strzero,\strone\}.
\end{align}
This is for $\mathcal{A}=\mathbb{N}$ and $\xi=2$ an example for the asynchronous 
$[\mathcal{A},\xi]$ channel with binary input as introduced in \cite{Yeung2009}.

The first way to argue that $\langle\mathcal{V},\mathcal{K}\rangle$ is not regular is to interpret the set of runs
\begin{align}
\{r|w(r)\in\mathcal{V},\,l(r)\in\mathcal{K}\}
\end{align}
as an infinite alphabet. The set $\langle\mathcal{V},\mathcal{K}\rangle$ is thus generated by an infinite alphabet, whereas a regular language is by definition generated by a finite alphabet \cite{Sipser2006}. However, the set $\langle\mathcal{V},\mathcal{K}\rangle$ can also be generated by concatenating the two runs $r_0,r_1$ with $w(r_0)=w(r_1)=1$ and $l(r_0)=\strzero$, $l(r_1)=\strone$. Indeed we now have a finite alphabet $\{r_0,r_1\}$ by which we can generate $\langle\mathcal{V},\mathcal{K}\rangle$, but we need an infinite memory: only the concatenation of $1,2,4,8,\dotsc$ runs of the same label result in a valid run-length. This memory cannot be implemented by a finite-state machine, which again shows that $\langle\mathcal{V},\mathcal{K}\rangle$ is not regular.

\subsection{How to Derive the Probabilistic Capacity of $\langle\mathcal{W},\mathcal{L}\rangle$}
Our aim is to derive the probabilistic capacity of $\langle\mathcal{W},\mathcal{L}\rangle$ in the general case, that is, we want to derive the maximum entropy rate an input process of $\langle\mathcal{W},\mathcal{L}\rangle$ can have. According to Theorem~\ref{theo:probabilisticCapacity}, the maximum entropy rate is upper bounded by the combinatorial capacity, which itself is, by Theorem~\ref{theo:combinatorialCapacity}, given by the abscissa of convergence of the generating function. Our approach is now as follows:
\begin{enumerate}
\item We first derive the generating function.
\item We then use the generating function to calculate the combinatorial capacity.
\item We finally define an input process whose entropy rate is equal to the combinatorial capacity.
\end{enumerate}
After accomplishing these three tasks, we have shown that the probabilistic capacity of $\langle\mathcal{W},\mathcal{L}\rangle$ is equal to its combinatorial capacity and that its value can be derived by using the corresponding formulas for the combinatorial capacity. It should be noted that this approach is also suitable for other types of constrained systems. For instance, it was used in \cite{Bocherer2010a} to derive the combinatorial and the probabilistic capacity of $(j,k)$ constraints.

%% file: generating.tex
\section{Generating Function of $\langle\mathcal{W},\mathcal{L}\rangle$}
\label{sec:generating}

We start by deriving the generating function of $\langle\mathcal{W},\mathcal{L}\rangle$. We do this run by run. The generating function of the run-length of the first run is given by 
\begin{align}
\gf_\mathcal{W}(s)=\sum\limits_{\nu\in\mathcal{W}}e^{-\nu s}.
\end{align}
For the first run, we can choose from among $|\mathcal{L}|$ labels, so the generating function of the first run $R_1$ is given by
\begin{align}
\gf_{R_1}(s)=|\mathcal{L}|\gf_\mathcal{W}(s).
\end{align}
The generating function of the run-lengths of the second run is again $\gf_\mathcal{W}(s)$, however, given the label chosen for the first run, we can for the second run only choose from among $|L|-1$ labels. The first two runs result from concatenating the first run with the second run. Since we guarantee that two subsequent runs have different labels, concatenating the first with the second run corresponds to multiplying the corresponding generating functions \cite[Chapter 2]{Bocherer2007a}. The generating function of the first two runs is thus
\begin{align}
\gf_{R_{2}}(s)=|\mathcal{L}|\gf_\mathcal{W}(s)(|\mathcal{L}|-1)\gf_\mathcal{W}(s).
\end{align}
The same as for the second run applies for all subsequent runs, so the generating function of the first $k$ runs is given by
\begin{align}
\gf_{R_{k}}(s)&=|\mathcal{L}|\gf_\mathcal{W}(s)\bigl[(|\mathcal{L}|-1)\gf_\mathcal{W}(s)\bigr]^{k-1}
\end{align}
To get the complete generating function of $\langle\mathcal{W},\mathcal{L}\rangle$ we have to add up all $\gf_{R_k}$.
\begin{align}
\gf_{\langle\mathcal{W},\mathcal{L}\rangle}(s)&=\sum_{k=1}^\infty \gf_{R_k}(s)\\
&=|\mathcal{L}|\gf_\mathcal{W}(s)\sum\limits_{m=0}^\infty\bigl[(|\mathcal{L}|-1)\gf_\mathcal{W}(s)\bigr]^m\label{eq:generatingFunctionWL}.
\end{align}

%% file: combinatorial.tex
\section{Combinatorial Capacity of $\langle\mathcal{W},\mathcal{L}\rangle$}
\label{sec:combinatorial}

With the help of the generating function \eqref{eq:generatingFunctionWL} as derived in the previous section, we can now derive the combinatorial capacity of $\langle\mathcal{W},\mathcal{L}\rangle$ by applying Theorem~\ref{theo:combinatorialCapacity}. Let $\repart(s)$ denote the real part of $s$. From \cite[Theorem 3]{Hardy1915}, we know that $\gf_{\langle\mathcal{W},\mathcal{L}\rangle}(s)$ converges if and only if $\gf_{\langle\mathcal{W},\mathcal{L}\rangle}[\repart(s)]$ converges, and since $\gf_{\langle\mathcal{W},\mathcal{L}\rangle}$ is strictly positive on the real axis, the latter converges if and only if
\begin{align}
(|\mathcal{L}|-1)\gf_\mathcal{W}[\repart(s)]<1.
\end{align}
Thus, as a corollary to Theorem~\ref{theo:combinatorialCapacity}, we have
\begin{proposition}
The combinatorial capacity $\mathsf{C}$ of  $\langle\mathcal{W},\mathcal{L}\rangle$ is given by the unique positive real solution of
\begin{align}
(|\mathcal{L}|-1)\gf_\mathcal{W}(s)=1.\label{eq:combinatorialCapacityRL}
\end{align}
\end{proposition}
This proposition coincides with \cite[Theorem 2]{Yeung2009}. As observed in \cite{Yeung2009}, for $|\mathcal{L}|=2$, the proposition reduces to \cite[Proposition 1.1]{Csiszar1969}. Furthermore, if $\mathcal{L}=\{0,1\}$  and $\mathcal{W}=\{1,2,\dotsc,k\}$, then the proposition provides the combinatorial capacity of the $(j=k,k)$ run-length constraint and coincides with the formulas derived in \cite{Mittelholzer2009,Bocherer2010a}.

%% file: probabilistic.tex
\section{Probabilistic Capacity of $\langle\mathcal{W},\mathcal{L}\rangle$}
\label{sec:probabilistic}
To calculate the probabilistic capacity of $\langle\mathcal{W},\mathcal{L}\rangle$, it remains to define an input process for $\langle\mathcal{W},\mathcal{L}\rangle$ that has an entropy rate equal to the combinatorial capacity $\mathsf{C}$.
\subsection{Preliminary Considerations}
\label{subsec:preliminaries}
 Not every process $Y=\{Y_i\}_{i=1}^\infty$ with $Y_i\in\langle\mathcal{W},\mathcal{L}\rangle$ is a valid input process. The first reason is that a concatenation of elements from $\langle\mathcal{W},\mathcal{L}\rangle$ is not necessarily an element of $\langle\mathcal{W},\mathcal{L}\rangle$. Let, for instance, $\langle\mathcal{W},\mathcal{L}\rangle$ be given by
\begin{align}
\mathcal{W} := \{1,2,\pi\},\quad
\mathcal{L} := \{\mathrm{red},\mathrm{green}\}.
\end{align}
Define the run $r$ by $w(r)=\pi$ and $l(r)=\mathrm{red}$. Run $r$ is an element of $\langle\mathcal{W},\mathcal{L}\rangle$. The string $\catop(r,r)$ is of color $\mathrm{red}$, therefore, $\catop(r,r)$ is just one run. But $w[\catop(r,r)]=2\pi\notin\mathcal{W}$, so the concatenation of $r$ with $r$ is not in $\langle\mathcal{W},\mathcal{L}\rangle$ and $Y$ violates condition 1) of Definition~\ref{def:inputProcess}. 

A second reason why $Y$ may not be an input process is ambiguity: define $r_1$ by $w(r_1)=1$ and $l(r_1)=\mathrm{red}$ and define $r_2$ by $w(r_2)=2$ and $l(r_2)=\mathrm{red}$. Obviously, $r_1,r_2\in\langle\mathcal{W},\mathcal{L}\rangle$. But $\catop(r_1,r_1)=r_2$, so $r_2$ can be a realization either of $(Y_1,Y_2)$ or just of $Y_1$. This violates condition 2) of Definition~\ref{def:inputProcess}. 

We leave the example and let from now on $\langle\mathcal{W},\mathcal{L}\rangle$ again be an arbitrary general run-length set with $|\mathcal{L}|\geq 2$ and $|\mathcal{W}|$ possibly countably-infinite. We choose the following approach to construct an input process that guarantees both condition 1) and condition 2) of Definition~\ref{def:inputProcess}: we define a subset $\mathcal{Y}$ of $\langle\mathcal{W},\mathcal{L}\rangle$ such that if the $Y_i$ take values in $\mathcal{Y}$, then both conditions from Definition~\ref{def:inputProcess} are automatically fulfilled.
\subsection{The Support of a Class of Input Processes of $\langle\mathcal{W},\mathcal{L}\rangle$}
We choose an arbitrary but fixed element $l_0\in\mathcal{L}$ and consider the subset $\mathcal{Y}$ of $\langle\mathcal{W},\mathcal{L}\rangle$ that consists of strings where the first run has label $l_0$ and where all subsequent runs have a label different from $l_0$. Note that each string in $\mathcal{Y}$ consists of at least two runs. The set $\mathcal{Y}$ has two important properties:
\begin{enumerate}
\item If $x\in\mathcal{Y}$ and $y\in\mathcal{Y}$ then $\catop(x,y)\in\langle\mathcal{W},\mathcal{L}\rangle$, so
\begin{align}
\bigcup_{i=1}^\infty \{\catop(x_1,\dotsc,x_i)|(x_1,\dotsc,x_i)\in\mathcal{Y}^i\} \subseteq \langle\mathcal{W},\mathcal{L}\rangle.\nonumber
\end{align}
\item Each string in $\langle\mathcal{W},\mathcal{L}\rangle$ that starts with a run of label $l_0$ and ends with a run of a label different from $l_0$ can be unambiguously generated by concatenating elements from $\mathcal{Y}$.
\end{enumerate}
In other words, any random process $\{Y_i\}_{i=1}^\infty$, $Y_i\in\mathcal{Y}$ fulfills condition 1) and condition 2) of Definition~\ref{def:inputProcess} and is therefore an input process of $\langle\mathcal{W},\mathcal{L}\rangle$.

In the following derivations, we will also need the generating function of $\mathcal{Y}$. It is given by
\begin{align}
\gf_\mathcal{Y}(s) &= 
\underbrace{\gf_\mathcal{W}(s)}_{\text{\parbox{1.3cm}{one run\\[-0.18cm]with label $l_0$}}}
\underbrace{(|\mathcal{L}|-1)\gf_\mathcal{W}(s)}_{\text{\parbox{2cm}{one run with label\\[-0.18cm]from $\mathcal{L}\setminus l_0$}}}
\underbrace{\sum\limits_{k=0}^\infty\Bigl[(|\mathcal{L}|-2)\gf_\mathcal{W}(s)\Bigr]^k}_{\text{\parbox{3cm}{$0$ or more runs with labels\\[-0.18cm]from $\mathcal{L}\setminus l_0$ but different from\\[-0.18cm]the label of the 2nd run}}}\label{eq:supportGf}.
\end{align}
The derivation of $\gf_\mathcal{Y}$ follows the lines of Section~\ref{sec:generating}.
\subsection{A Maxentropic Input Process of $\langle\mathcal{W},\mathcal{L}\rangle$}
For a random process that takes values in $\mathcal{Y}$ as defined in the previous subsection, we now want to define a PMF in such a way that the entropy rate of the resulting input process is equal to the combinatorial capacity $\mathsf{C}$ and thereby maxentropic. To this end, we will need the following proposition:
\begin{proposition}
\label{prop:entropyRate}
Let $Z=\{Z_i\}_{i=1}^\infty$ denote a random process with each $Z_i$ taking values in the countably infinite set $\mathcal{Z}$ and let $w\colon \mathcal{Z}\rightarrow\mathbb{R}_{>0}$ be an associated positive weight function. Let $R$ denote the unique positive real solution of
\begin{align}
\gf_\mathcal{Z}(s)=1
\label{eq:maximumEntropie}
\end{align}
and define $p_Z$ as
\begin{align}
p_{Z}(z):=e^{-w(z)R},\quad z\in\mathcal{Z}.
\end{align}
Then $p_Z$ is a PMF and by letting the $Z_i$ be independent and identically distributed (IID) according to $p_Z$,
\begin{align}
\bar{\entop}(Z)=R.
\end{align}
\end{proposition}
\begin{IEEEproof}
For a finite set $\mathcal{Z}$, this proposition is equivalent to \cite[Theorem 1]{Krause1962}. By \cite[Lemma 1]{Bocherer2010a} from our work, the generalization to countably infinite $\mathcal{Z}$ follows directly.
\end{IEEEproof}
Let $Y=\{Y_i\}_{i=1}^\infty$ denote a random process where the $Y_i$ take values in $\mathcal{Y}$, with $\mathcal{Y}$ as defined in the previous subsection. We first show that $\gf_\mathcal{Y}(\mathsf{C})=1$. Recall from \eqref{eq:combinatorialCapacityRL} that 
\begin{align}
(|\mathcal{L}|-1)\gf_\mathcal{W}(\mathsf{C})=1.\label{eq:equalOne}
\end{align}
We thus have
\begin{align}
\gf_\mathcal{Y}(\mathsf{C})
&= \gf_\mathcal{W}(\mathsf{C})\underbrace{(|\mathcal{L}|-1)\gf_\mathcal{W}(\mathsf{C})}_{=1}\sum\limits_{k=0}^\infty\Bigl[\underbrace{(|\mathcal{L}|-2)\gf_\mathcal{W}(\mathsf{C})}_{=1-1\cdot\gf_\mathcal{W}(\mathsf{C})}\Bigr]^k\label{eq:der:supportGf}\\
&=\gf_\mathcal{W}(\mathsf{C})\sum\limits_{k=0}^\infty\bigl[1-\gf_\mathcal{W}(\mathsf{C})\bigr]^k\label{eq:der:equalOne}\\
&=\frac{\gf_\mathcal{W}(\mathsf{C})}{1-[1-\gf_\mathcal{W}(\mathsf{C})]}\label{eq:der:geometricSeries}\\
&=\frac{\gf_\mathcal{W}(\mathsf{C})}{\gf_\mathcal{W}(\mathsf{C})}=1
\end{align}
where \eqref{eq:der:supportGf} follows from \eqref{eq:supportGf}, \eqref{eq:der:equalOne} follows from \eqref{eq:equalOne}, and where we used the geometric series formula in \eqref{eq:der:geometricSeries}. By Proposition~\ref{prop:entropyRate},
\begin{align}
p_{Y}(y):=e^{-w(y)\mathsf{C}},\quad y\in\mathcal{Y}
\end{align}
is a PMF. We let the $Y_i$ be IID according to $p_{Y}$. As a consequence, again by Proposition~\ref{prop:entropyRate},
\begin{align}
\bar{\entop}(Y)&=\mathsf{C}\label{eq:der:bound}
\end{align}
Thus, the entropy rate of $Y$ is equal to the combinatorial capacity $\mathsf{C}$, and since $\mathsf{C}$ is according to Theorem~\ref{theo:probabilisticCapacity} an upper bound, $Y$ is maxentropic. We thus have the following proposition.
\begin{proposition}
\label{prop:probabilisticCapacityRL}
The probabilistic capacity of $\langle\mathcal{W},\mathcal{L}\rangle$ is equal to its combinatorial capacity.
\end{proposition}

%% file: discussion.tex
\subsection{Discussion of \cite{Yeung2009}}
\label{sec:discussion}
\subsubsection{Definition of probabilistic capacity}
Proposition~\ref{prop:probabilisticCapacityRL} was proved in \cite{Yeung2009} with respect to the following definition of probabilistic capacity. Let $Y=\{Y_i\}_{i=1}^\infty$ denote an input process of $\langle\mathcal{W},\mathcal{L}\rangle$. Let $Y$ generate strings until time instant $T$. If the last run is not complete, discard the incomplete last run. Let $\nu_0$ denote the smallest element from $\mathcal{W}$. The strings that $Y$ can generate in this way are for each time instant $T$ given by the set
\begin{align}
\mathcal{X}_T=\{s\in\langle\mathcal{W},\mathcal{L}\rangle|w(s)\leq T,\,w(s)+\nu_0>T\}.
\end{align}
Define for each time instant $T$ an auxiliary random variable $X_T$ with the support $\mathcal{X}_T$. Since the input process $Y$ generates one of the strings from $\mathcal{X}_T$ until time instant $T$ with probability $1$, the PMF of $Y$ implies a PMF of $X_T$. The authors of \cite{Yeung2009} now define the entropy rate of $Y$ by
\begin{align}
\bar{\entop}(Y):=\limsup_{T\rightarrow\infty} \frac{\entop(X_T)}{T}\label{eq:entropyRateYeung}
\end{align}
and define the probabilistic capacity of $\langle\mathcal{W},\mathcal{L}\rangle$ by the maximum of $\bar{\entop}$ over all input processes $Y$. With respect to this definition, it is easy to show Theorem~\ref{theo:probabilisticCapacity}, but it is difficult to show Proposition~\ref{prop:probabilisticCapacityRL}. This observation is reflected by the length of the proof of \cite[Theorem 4]{Yeung2009} given in \cite{Yeung2009}. As a side-result of our work, both definitions of probabilistic capacity are equivalent, but their calculations differ.
\subsubsection{Maxentropic input process}
In \cite{Yeung2009}, the authors solve the problem of violating the conditions from Definition~\ref{def:inputProcess} by defining an input process with memory, specifically, a Markov chain with $|\mathcal{L}|$ states. Basically, they consider PMFs of $Y$ where 
\begin{align}
\probop[l(Y_{i+1})=\mathtt{t}|l(Y_{i})=\mathtt{t}]=0,\quad\forall\mathtt{t}\in\mathcal{L}
\end{align}
i.e., two subsequent runs generated by the process have the same label with probability zero. It can be shown that the entropy rate as defined in Definition~\ref{def:entropyRate} of this process is equal to the combinatorial capacity $\mathsf{C}$, i.e., this process is also maxentropic with respect to our definition of entropy rate.

%% file: conclusions.tex
\section{Conclusions}

In this work, we used generating functions to show in a unified manner that the probabilistic capacity of a general run-length set is equal to its combinatorial capacity. This is an interesting result, since general run-length sets are in general not regular and can not be analyzed within the framework of finite state machines, which has been the usual approach for the analysis of constrained systems so far. Generating functions and their information-theoretic properties, which we recently established, appear to be a strong tool for the analysis of a broad class of constrained systems. We believe that known results can be re-established within this new framework in a unified manner and that new results can be obtained. The present work is a first step in this direction.